\pdfoutput=1
\documentclass[amsmath,amssymb,prl,hyperlink,twocolumn]{revtex4}

\usepackage{graphicx}
\usepackage{soul}
\usepackage[colorlinks=true,citecolor=blue,linkcolor=magenta]{hyperref}
\usepackage[usenames]{color}
\usepackage{amsfonts}
\usepackage{times}

\begin{document}

\title{Dynamically Controlled Resonance Fluorescence from a Doubly Dressed Solid-State Single Emitter}

\author{Yu He$^{1,2}$, Y.-M. He$^{1,2}$, J. Liu$^{3,4}$,  Y.-J. Wei$^{1,2}$, H. Ramirez$^{1,2}$, M. Atat\"{u}re$^{5,1}$, C. Schneider$^6$, M. Kamp$^6$, \\ S. H\"{o}fling$^{6,1,2,7}$, C.-Y. Lu$^{1,2}$, J.-W. Pan$^{1,2}$ \vspace{0.2cm}}

\affiliation{$^1$ Hefei National Laboratory for Physical Sciences at Microscale and Department of Modern Physics, University of Science and Technology of China, Hefei, Anhui 230026, China}
\affiliation{$^2$ CAS Centre for Excellence and Synergetic Innovation Centre in Quantum Information and Quantum Physics, University of Science and Technology of China, Hefei, Anhui 230026, China}
\affiliation{$^3$ Institute for Research in Electronics and Applied Physics, University of Maryland, College Park, MD 20742, USA}
\affiliation{$^4$ South China Academy of Advanced Optoelectronics, South China Normal University, Guangdong, Guangzhou 510006 China}
\affiliation{$^5$ Cavendish Laboratory, JJ Thomson Avenue, University of Cambridge, CB3 0HE Cambridge, UK}
\affiliation{$^6$ Technische Physik, Physikalisches Instit\"{a}t and Wilhelm Conrad R\"{o}ntgen-Center for Complex Material Systems, Universitat W\"{u}rzburg, Am Hubland, D-97074 W\"{u}zburg, Germany}
\affiliation{$^7$ SUPA, School of Physics and Astronomy, University of St. Andrews, St. Andrews KY16 9SS, United Kingdom}

\date{\vspace{0.1cm}\today}

\begin{abstract}
We report the first experimental demonstration of interference-induced spectral line elimination predicted by Zhu and Scully [Phys. Rev. Lett. \textbf{76}, 388 (1996)] and Ficek and Rudolph [Phys. Rev. A \textbf{60}, 4245 (1999)]. We drive an exciton transition of a self-assembled quantum dot in order to realize a two-level system exposed to bichromatic laser field and observe nearly complete elimination of the resonance fluorescence spectral line at the driving laser frequency. This is caused by quantum interference between coupled transitions among the doubly dressed excitonic states, without population trapping. We also demonstrate multiphoton ac Stark effect with shifted subharmonic resonances and dynamical modifications of resonance fluorescence spectra by using double dressing.
\end{abstract}


\maketitle

Self-assembled semiconductor quantum dots (QDs) have attracted considerable interest in both fundamental studies of quantum optics and promising applications in optoelectronics and quantum information \cite{Shields07,Warburton13}, as their atom-like properties can be tailored artificially through epitaxial growth and band-gap engineering. Their compatibility with semiconductor nanofabrication technologies enables monolithical integration to micro- and nano-structures to enhance light-matter interaction \cite{PeterLodahl13,Jahnke12}. The atom-like behaviors of QDs have been confirmed by numerous experiments such as photon antibunching \cite{Michler00,Santori01} and spontaneous emission control \cite{Wang11,Englund05}.  More recently, coherent control of the QD two-level systems has allowed the observations of Rabi oscillation \cite{Ramsay10}, Ramsey interference \cite{Jayakumar13}, Autler-Townes splitting \cite{Xu07}, Mollow absorption \cite{Xu08PRL}, and resonance fluorescence triplet \cite{Muller07RF_QD,Vamivakas09QDRF}, which can be well preserved even under strong optical excitations. Resonance fluorescence from QDs has also provided a versatile tool in spectroscopy \cite{Flagg09}, spin readout \cite{Vamivakas10} and high-quality single-photon source \cite{Ates09,Yuming13}.

New physics beyond the Mollow fluorescence triplet arises when a two-level system is exposed to strong bichromatic driving fields, exhibiting a range of nonlinear and multiphoton dynamics of light-matter interaction \cite{Rudolph98Subharmonic}. In particular, it has been pointed out that a dynamical cancellation of the spontaneous emission spectral line, initially predicted for a three-level system \cite{ShiYao96}, may be possible in a two-level atom \cite{Ficek99DoubleDress,Ficek96DoubleOurM}, which is not due to population trapping, but induced by quantum interferences between different transition channels in a doubly dressed atom. This occurs when a two-level atom is driven by a strong resonant laser with a Rabi frequency $2\Omega$ \cite{RabiDefinition}, and a second, weaker laser detuned by $2\Omega$ from the strong field. The suppression of spontaneous emission spectral lines also appears when the second laser is coupled to the shifted subharmonics of $2\Omega$, which can be attributed to the multiphoton ac Stark effect \cite{Rudolph98Subharmonic}. Despite the extensive efforts undertaken to observe this phenomenon \cite{Xia96SGC,LiLi00Comment,Berman98Xia_Comment} the experimental demonstration of a complete cancellation of spontaneous emission lines has been claimed once in a V-type level atom \cite{Xia96SGC}, where was also an alternative interpretation based on population trapping possible \cite{Berman98Xia_Comment}. Here, we use a single QD, doubly dressed by bichromatic fields to demonstrate unequivocally the complete cancellation of spontaneous emission spectral lines and other features arising from this phenomenon including shifted subharmonic resonances and modifications of sideband spectra.

\begin{figure*}[htbp]
\includegraphics[width=1.00\textwidth,angle=0]{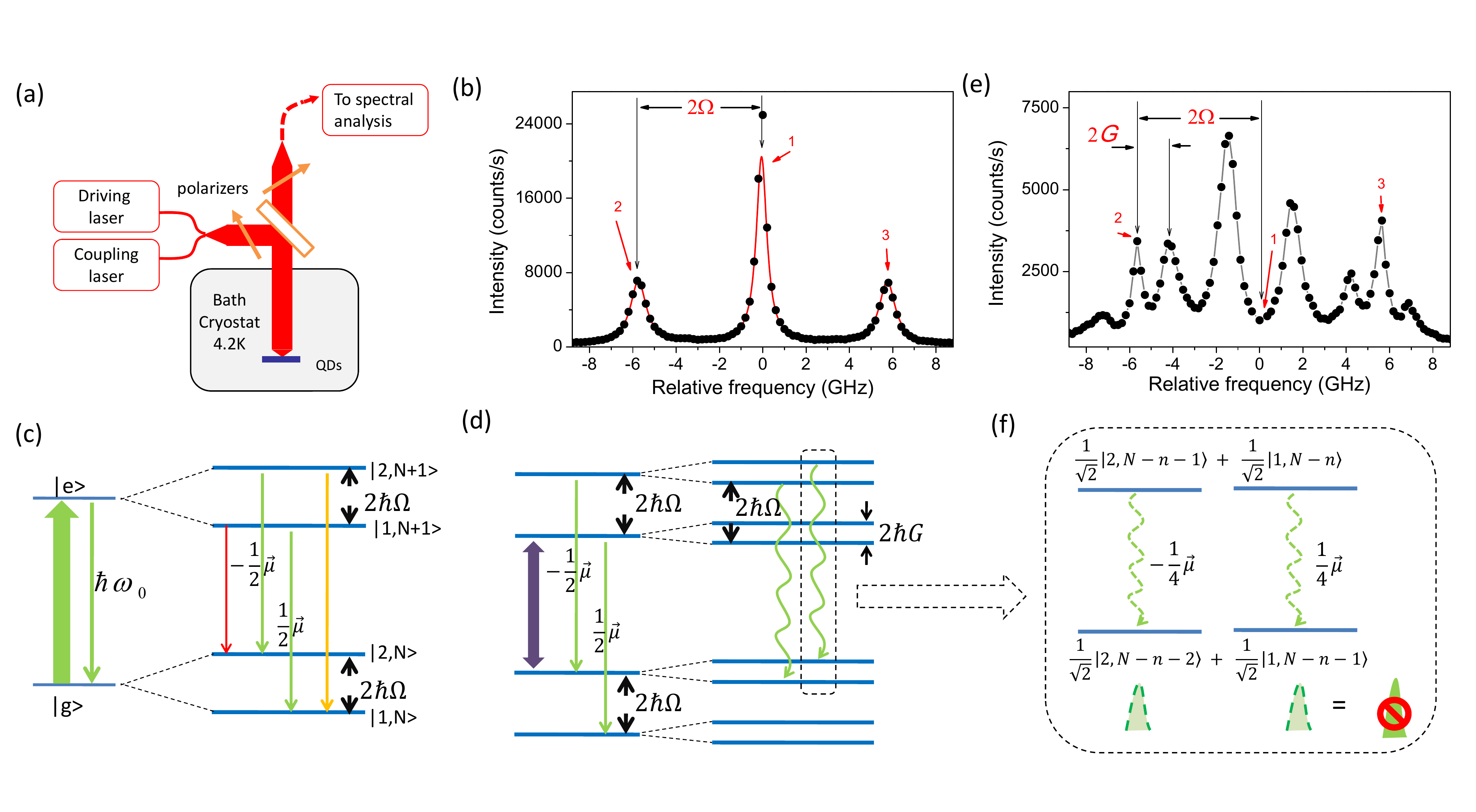}
\caption{\label{Fig.1} Resonance fluorescence: from Mollow triplet to the doubly dressed spectrum. (a) Schematic of the experiment setup. (b) Dressed QD spectrum showing a typical Mollow triplet. The splitting is Rabi frequency of the driving laser $2\Omega=5.8\pm0.1$ GHz. Black circles are the experimental data and the red curve is the theoretical fit. (c) Energy levels of singly dressed states of the QD excited by a strong resonant pumping laser (thick green arrow). The ground state $|g\rangle$ and excited state $|e\rangle$ of the QD are split into doublet states with a splitting energy $2\hbar\Omega$. The two transitions related to the central peak of Mollow triplet are anti-parallel (green lines), but decay with no correlation. The yellow and red lines correspond to two side-peak transitions. (d) Energy levels of doubly dressed states of the QD by introducing a coupling laser (thick purple arrow) resonant to the inner doublet transition. Every singly dressed state splits into new doublets with a spacing of $2\hbar G$, where $2 G$ is the coupling field Rabi frequency. The transitions corresponding to the central peak in doubly dressed spectrum are represented in green wavy arrows. (e) Doubly dressed QD spectrum. The central peak is strongly suppressed. The splitting of the peaks within the daughter triplet is $2 G= 0.6\Omega$. (f) Tracing transition , from $|N+M,n-\rangle$ to $|N+M-1,n-\rangle$, back to their eigenstates. The selection-rules-allowed transitions are indicated by dashed wavy arrows. These two transitions possess opposite phases and destructively interfere, which leads to the suppression of the spontaneous emission at the driving laser frequency (For simplicity, weak field photon number M is abbreviated and not shown).}
\end{figure*}

Our experiments are carried out on self-assembled InGaAs QDs embeded in a low-$\mathrm{Q}$ planar microcavity ($\mathrm{Q} \approx 200$) and housed in a cryogen-free bath cryostat operating at 4.2 K, as shown in Fig.~1(a). The measurements are performed on the neutral exciton transition of a single QD with a confocal microscope allowing bichromatic excitations of the QD from the side arm and collection of the emitted fluorescence through the top arm. By using a cross-polarization configuration the resonance fluorescence is separated from the excitation lasers with a signal to background ratio exceeding 28:1 for the highest excitation power of 100 $\mu$W ($2\Omega\approx14.2\gamma_{sp}$) used in this experiment, in which $\gamma_{sp}$ denotes spontaneous emission rate of the neutral exciton state.

Figure 1(b) shows a typical Mollow triplet spectrum from a single QD resonantly driven by a strong, narrow-band ($<$2 MHz) laser. Exciton states in QDs are treated as two-level systems with ground state $|g\rangle$, excited state $|e\rangle$, transition frequency $\omega_{0}$ and dipole transition moment $\vec{\mu}$. Due to the light-matter interaction induced by the pumping laser, the dressed states (doublets) are formed:
\begin{eqnarray}
| i,N \rangle =\frac{1}{\sqrt2}[| g,N \rangle-(-1)^{i}| e,N-1 \rangle], i=1,2
\end{eqnarray}
with energies $E_{i,N}=\hbar(N\omega_{0}\pm\Omega)$, where $N$ is the number of photons in the driving field and $2\Omega$ is the Rabi frequency. The four transitions between neighboring doublets result in three distinct transition frequencies $\omega_{0}$ and $\omega_{0}\pm2\Omega$ in the Mollow triplet spectrum, see Fig.~1(c). Exciton lifetime $T_{1} = 390$ ps is obtained from time-resolve measurements while Rabi frequency of the driving laser $2\Omega=5.8\pm0.1$ GHz and coherence time of the exciton state $T_{2} = 424$ ps are extracted by fitting the Mollow triplet spectrum (see SI 2). It is worthwhile to notice that the two transitions contributing to the central peak exhibit anti-parallel dipole moments, i.e., $\pm\frac{1}{2} \vec{\mu}$, but the calculated correlation functions of these transitions $\langle\sigma^{+}_{iiN}\sigma^{-}_{jjN}\rangle(i\ne j)$ are equal to zero \cite{Ficek99DoubleDress}, indicating these two transitions decay independently. $\sigma^{\pm}_{iiN}=| i,N \rangle\langle N-1,j|$ ($i,j=1,2$) is transition dipole moment operator of the dressed atom transition.

\begin{figure*}[htbp]
\includegraphics[width=1.00\textwidth,angle=0]{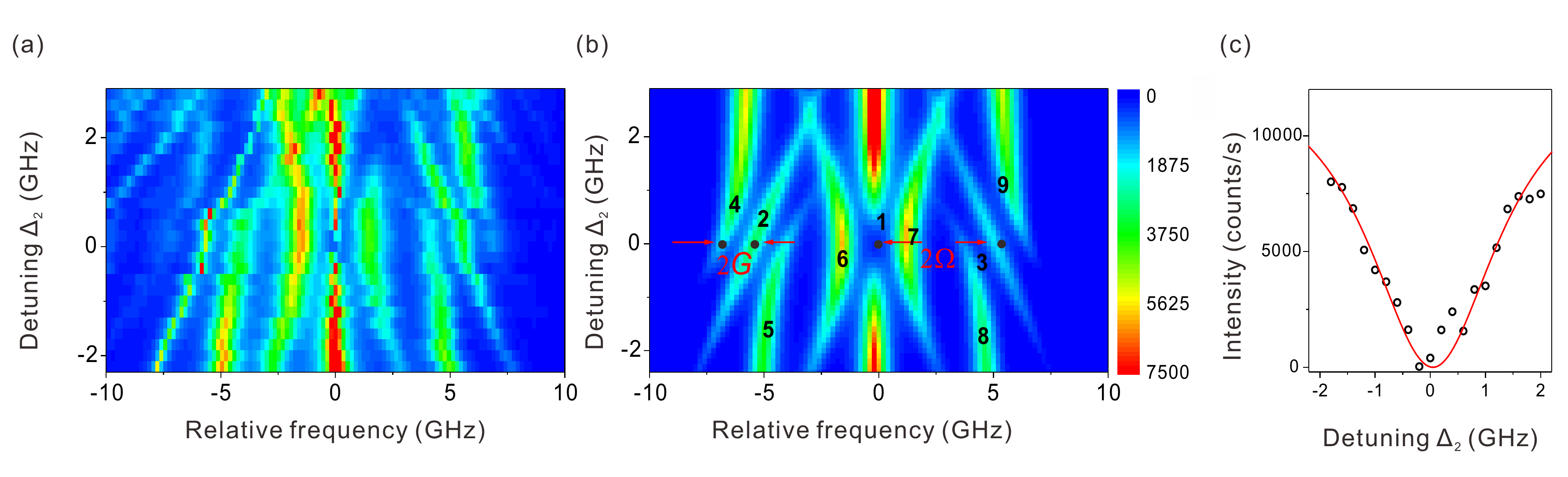}
\caption{\label{Fig2.}Spectra of the doubly dressed QD.
(a) Measured spectra of the doubly dressed QD as a function of detuning $\Delta_{2}$. (b) The simulation of doubly dressed spectra. Driving laser Rabi frequency $2\Omega$ associates to the splitting of original Mollow triplet peak 1,2 and 1,3 while the coupling laser Rabi frequency $2G$ relates to the inner splitting of the peaks in the daughter triplet. (a), (b) share the same color bar. (c) The central peak intensity versus the coupling laser detuning. Counts at zero detuning go to dark counts level. Black circles are the experimental data and the red curve is the theoretical fit.
}
\end{figure*}

We then introduce a second coupling laser with frequency $\omega_{2}=\omega_{0}-2\Omega$ and Rabi frequency $2 G=0.6\Omega$ to couple the degenerate transitions that relate to the central peak, as shown in Fig.~1(d). The newly formed doubly dressed states can be written as $|N+M,n\pm\rangle=\frac{1}{\sqrt2}[|2,N-n-1,M+n+1\rangle\pm|1,N-n,M+n\rangle]$, where $N$, $M$ are photon number of the driving field and the coupling field respectively and \textit{n} is the sub-level number within the doublets. Every spectral emission line in the Mollow triplet splits into a daughter triplet because of the formation of quartet states by dressing the doublet states, as shown in Fig.~1(d). In such a spectrum 9 emission lines in total are expected, however, in the measured spectrum shown in Fig.~1(e) the central emission line is completely suppressed as a result of the destructive quantum interference induced by the coupling laser.

This quantum interference is easier to interpret by tracing the transitions between doubly dressed states back to their eigenstates. One of the transitions corresponding to the central peak, from $|N+M,n+\rangle$ to $|N+M-1,n+\rangle$, is highlighted in Fig.~1(d) and traced back to the superimpositions of their eigenstates (singly dressed states) in Fig.~1(f):  $|N+M,n+\rangle=\frac{1}{\sqrt2}[|2,N-n-1,M+n+1\rangle+|1,N-n,M+n\rangle]$, $|N+M-1,n+\rangle=\frac{1}{\sqrt2}[|2,N-n-2,M+n\rangle+|1,N-n-1,M+n-1\rangle]$. Among the possible decay channels between these eigenstates, only the transitions from $|2,N-n-1,M+n+1\rangle$ to $|2,N-n-2,M+n\rangle$ and from $|1,N-n,M+n\rangle$ to $|1,N-n-1,M+n\rangle$ are allowed by selection rules. Detailed calculations (see SI 4) reveal that these two transitions should be anti-phased and interfere destructively, eliminating the central spectral line. Similarly, other degenerate transitions corresponding to the central peak e.g., transition from $|N+M,n-\rangle$ to $|N+M-1,n-\rangle$, also exhibit destructive interferences.

A quantitative comparison between the predicted and the measured spectra is performed by scanning the coupling laser across the resonance dip ($\omega_{0}-2\Omega$). Frequency detuning between the coupling laser and inner transition of the doublets is denoted as $\Delta_{2}=\omega_{2}-(\omega_{0}-2\Omega)$. Fig.~2(a) represents a 2D contour plot of $\Delta_{2}$-dependent doubly dressed spectra which are well reproduced in the simulation of Fig.~2(b) by using the parameters extracted from the experiments (see SI 5[A]). The driving laser detuning $\Delta_{1}=-0.977$ GHz used in the simulation is obtained from a fit of the central peak intensity as a function of $\Delta_{2}$ in Fig.~2(a), as shown in Fig.~2(c) (See SI 5[B]). Comparing the experimental spectra to the prediction, we find both the peak evolution trends and their corresponding strengthes agree remarkably well. The anti-crossing feature of peak 4, 5 (6, 7 or 8, 9) in Fig.~2(b) is an indication of the coupling laser induced interaction while peaks 2,3 evolve linearly with coupling laser detuning $\Delta_{2}$. As the coupling laser moves away from the resonance, the centre peak suppressed by quantum interference gradually emerges and the Mollow triplet recovers at large detuning ranges.

To investigate the multiphoton absorption process and spectral eliminations at subharmonic resonances, we tune the coupling laser from the bare exciton energy $\omega_{0}$ to the inner doublet transition $\omega_{0}-2\Omega$, i.e., the frequency difference between the coupling laser and the bare exciton energy $\Delta_{3}=\omega_{2}-\omega_{0}$ is tuned from zero to -$2\Omega$. A 140-MHz etalon with 9.18 GHz free spectral range is employed to filter out the central peak fluorescence signal when tuning the coupling laser. The centre peak intensity as a function of $-\Delta_{3}$ is plotted in Fig.~3 and spectral line suppressions are clearly observed near to the subharmonic Rabi frequencies $2\Omega/n$ ($n=1\sim5$). Instead of quantum interference induced by single-photons, here the coupling laser via a n-photon process \cite{Mossberg01Multiphoton} couples the dressed states created by the strong driving field, which results in a dynamic Stark shift of the states and effectively shifts the positions of the subharmonics. The shifts of the subharmonics can be calculated in the framework of high-order perturbation theory by taking all other doublets states within the same energy manifold into account \cite{Rudolph98Subharmonic} as $\Delta_{1}=(1/8){\alpha}^2\Omega$ for $n=1$ and $\Delta_{n}=n{\alpha}^2\Omega/[2(n^{2}-1)]$ for n$>$1, where $\alpha$ is the power ratio between the coupling and driving laser. The calculated frequency displacements are 0.13 GHz, 0.34 GHz, 0.19 GHz for $2\Omega$, $2\Omega/2$, $2\Omega/3$, respectively and are very close to the shifts 0.11 GHz, 0.24 GHz, 0.20 GHz observed in the experiments. The incomplete suppression of the centre peak at subharmonics is an effect of the reduced splitting of the doubly dressed states in higher multiphoton processes \cite{Rudolph98Subharmonic}. The spectral cancellation induced by quantum interferences is different from the processes with population trapped in a specific state, such as coherent population trapping \cite{CPT} or electromagnetically induced transparency \cite{MuckeRempe10EITatom}. In our system, population $P_{n\pm,\overline{N}}=\langle\pm{n},\overline{N}|\rho|\overline{N},n\pm\rangle$ is evenly distributed within these states \cite{Ficek99DoubleDress}, where $|\overline{N},n\pm\rangle$ is doubly dressed eigenstate, $\overline{N}=M+N$ is total photon number, and $\rho$ is state density matrix.

\begin{figure}[tbp]
\includegraphics[width=0.50\textwidth,angle=0]{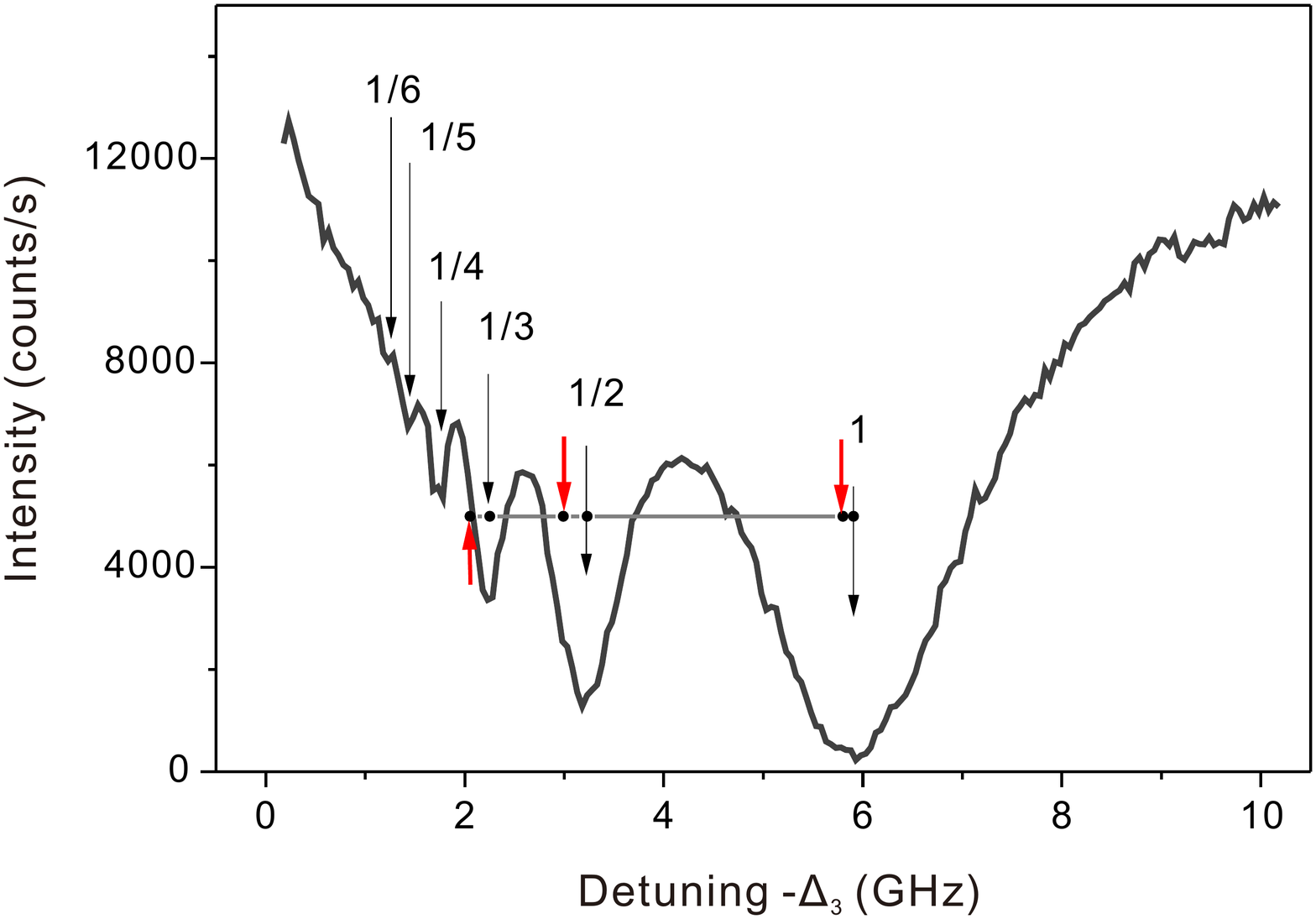}
\caption{\label{Fig3.}Subharmonics of the central peak for the doubly dressed QD. The positions of the subharmonics observed in experiments are indicated by black arrows while the un-shifted frequencies for the first three subharmonics are shown by red arrows, exhibiting observable shifts as a result of multiphoton ac Stark shift.
}
\end{figure}

Finally, the coupling laser is tuned to zero detuning with the exciton energy, i.e., exactly the same frequency as the driving laser. This degenerate excitation scheme generates a modified Mollow triplet in which the two side-peaks evolve into flat and broad sidebands, as predicted by Freedhoff and Ficek \cite{Freehoff97ATspectraM}. Panels a and c of Fig.~4 represent the experimental spectra for the power ratios of $\alpha=0.2$ and $\alpha=0.4$ respectively while the simulated spectra are shown in Fig.~4(b,d). The plateau of the sidebands originates from the continuum energy levels with the same population when dressing the dressed QDs at the same frequency.

Here we give a concise physical picture of this process as schematically illustrated in Figs.~4(e)-(g). Firstly, given the frequency degeneracy of the two excitation lasers, the doublets extend to degenerate doublet manifolds and each energy level has infinite inner states before additional effects due to the coupling laser, see Fig.~4(e). Secondly, owing to the interaction between the weak field and singly dressed states, infinite inner states shift from each other, forming a band of continuum states. Each pair of the lifted inner doublet sub-states shows a different energy splitting consequently exhibits a Mollow triplet spectrum with side-peaks shifted but central peak remained, see Fig.~4(f). These spectra convolve to create the spectrum shown in Fig.~4(g). This effect might be of importance in simulating continuum energy levels for studying light-matter interaction. Moreover, it can be used directly to develop a tunable wide-band single-photon sources.

\begin{figure*}[tbp]
\includegraphics[width=1.0\textwidth,angle=0]{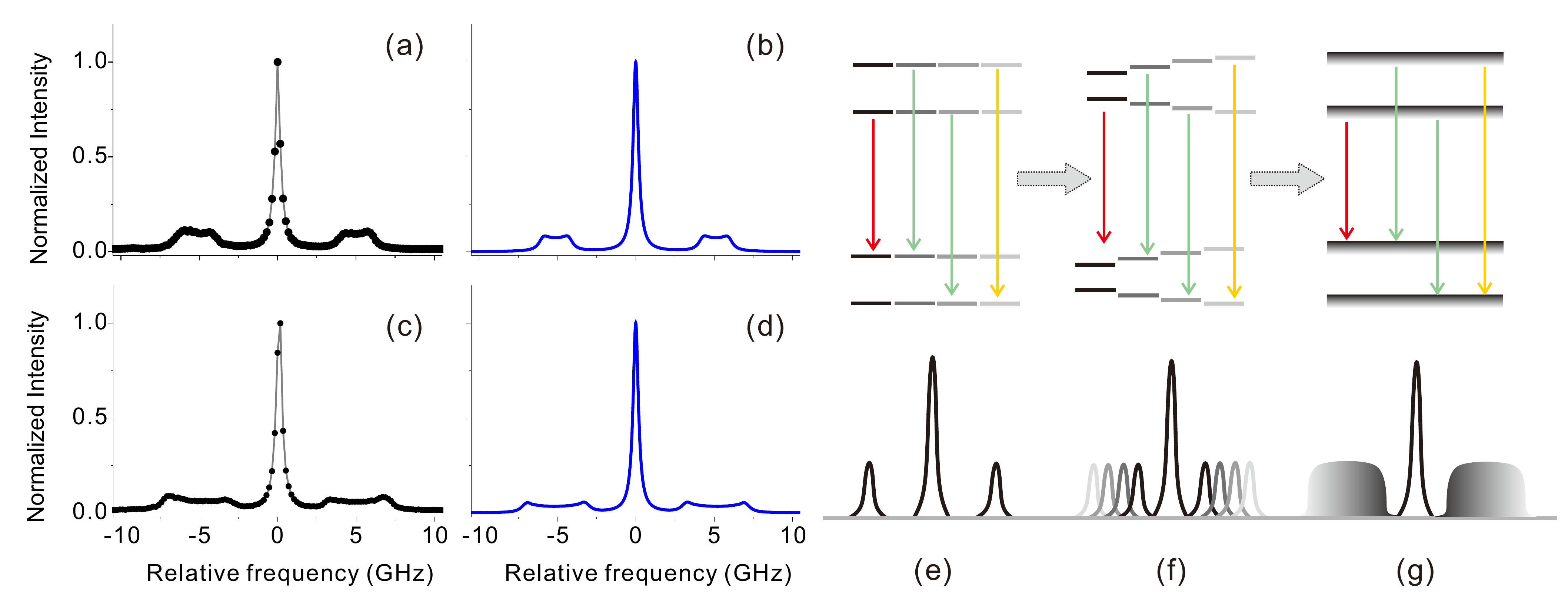}
\caption{\label{Fig4.}Sideband modulation by doubly dressing the QD. (a) and (b) are experimental and simulated spectra for $\alpha$=0.2 respectively. (c) and (d) are experimental and simulated spectra for $\alpha$=0.4 respectively. The x-axis and y-axis are the same in (a) - (d). (e) Energy levels of doubly dressed QDs without considering the interaction with the coupling laser. (f) Degeneracy lifting of the doubly dressed states by taking the coupling laser into account. (g) Continuum energy levels and flat sidebands of the QD driven by two lasers with the same frequency.
}
\end{figure*}

In conclusion, we have for the first time demonstrated spontaneously induced spectral cancellation together with multiphoton ac Stark shift and modulations of resonance fluorescence spectra using the transitions of a solid-state quantum emitter. Our results indicate that the self-assembled QD system accompanying with its advantages of stability, near background free fluorescence, long lifetime and short linewidth, serves a complementary test-bed to natural atoms for quantum physics research. The bichromatic excitation scheme provides a unique and versatile tool to control and modulate the spontaneous emission process especially in vacuum where modifications of the photonic density of states are not feasible. By further exploring the quantum interference generated by spontaneous emission, spectral line narrowing \cite{PengZhou96Narrow}, lasing without inversion \cite{PengZhou97NarrowTransGain} and quantum entanglement generation \cite{Tang10Entanglement_SGC} can be envisioned in QDs. Along with this direction, our work may facilitate the study of complicated spectra of two-level systems under multi-color laser excitations and stimulate the research on novel quantum interference phenomena in solid-state quantum systems.

\vspace{0.1cm}
\textit{Acknowledgement}: We acknowledge P. Wu, and M. Cramer for useful discussions. This work was supported by the National Natural Science Foundation of China, the Chinese Academy of Sciences and the National Fundamental Research Program. This work was also supported by the Alexander von Humboldt Foundation, the ERC. Y.H. and Y.-M.H. contributed equally to the work.


\begin{thebibliography}{10}
\bibitem{RabiDefinition}
The Rabi frequency here is defined as 2$\Omega$, which follows the definition in references \cite{Ficek99DoubleDress} and \cite{Ficek96DoubleOurM}.
\end{thebibliography}
\end{document}